*Article*

# A Tale of Two Entangled Instabilities—The Dual Role of *δ-O* in *HgBa$_2$Ca$_{n-1}$Cu$_n$O$_{2(n+1)+\delta}$*

**Itai Panas**

Division of Energy and Materials, Department of Chemistry and Chemical Engineering,
Chalmers Uniersity of Technology, Kemivägen 4, S-41296 Gothenburg, Sweden;
E-Mail: itai@chalmers.se; Tel.: +46-31-772-2860; Fax: +46-31-772-2853

**Abstract:** Low-energy instabilities in the hole-doped cuprates include, besides short range antiferromagnetic fluctuations and superconductivity, also ubiquitous translational and rotational symmetry breakings. The overwhelming majority of interpretations of these possibly related properties rely on mappings onto three bands spanned by the three atomic orbitals Cu3d($x^2$–$y^2$)($\sigma$), O2p$_x$($\sigma$), and O2p$_y$($\sigma$), these three local orbitals spanning the Zhang–Rice band (ZRB), the lower Hubbard bands (LHB) and the upper Hubbard bands (UHB), respectively. Here we demonstrate by means of supercell Density Functional Theory (DFT) (a) how oxygen intercalation affects the structures of the buffer layers, and (b) how the attenuated crystal field pulls two additional oxygen bands in the CuO$_2$ plane to the Fermi level. The self-consistent changes in electronic structure reflected in the corresponding changes in external potential comprise formal properties of the Hohenberg–Kohn theorems. Validation of present days' approximate exchange-correlation potentials to capture these qualitative effects by means of supercell DFT is made by comparing computed doping dependent structural shifts to corresponding experimentally observed correlations. The simplest generalization of Bardeen–Cooper–Schrieffer (BCS) theory is offered to articulate high-critical temperature superconductivity (HTS) from a normal state where crystal field causes states related to two non-hybridizing bands to coalesce at E$_F$.

## 1. Introduction

Holes doping of the CuO$_2$ planes in the HgBa$_2$CuO$_{4+\delta}$ superconductors, a representative of the cuprate class of superconductors discoverer three decades ago [1], is accomplished by intercalation of electrons accepting oxygen into the so-called charge reservoir [2]. While affecting the structure of these buffer layers the sole role of the added oxygen relevant to high-critical temperature superconductivity (HTS) is commonly assumed to comprise control the charge carrier concentration in the Zhang–Rice band (ZRB) [3], lower Hubbard bands (LHB), and upper Hubbard bands (UHB) manifold residing in the CuO$_2$ plane. Any symmetry breaking instability relevant to HTS is taken to be an innate property of these bands, albeit a function of the holes concentration [4–7]. Here we provide evidence for the surfacing of additional bands in HgBa$_2$CuO$_{4+\delta}$ that join the bands of local σ symmetry already present at $E_F$ upon oxygen doping. This occurs due to the system's sensitivity to changes in local crystal field [8]. In non-doped HgBa$_2$CuO$_4$ the low-dispersive bands of O2p(π) origin reside 0.5–1 eV below $E_F$. Deliberate shift of these bands to $E_F$ can be accomplished by displacing the A site ions away from the adjacent CuO$_2$ planes as was done for Bi2201 [9]. This is repeated in Figure 1 for undoped HgBa$_2$CuO$_4$ for 0 Å, ±0.1 Å, and ±0.2 Å displacements of Ba ions away from the CuO$_2$ plane. The reason for the surfacing of O2p(π) band is that its stability relies crucially on the Madelung potential which may become attenuated as the buffer layer relaxes owing to the doping. It is reminded here that while O displays an electron affinity of 1.4 eV in vacuum, the electron affinity of O¯ in vacuum is negative by ~ 10 eV.

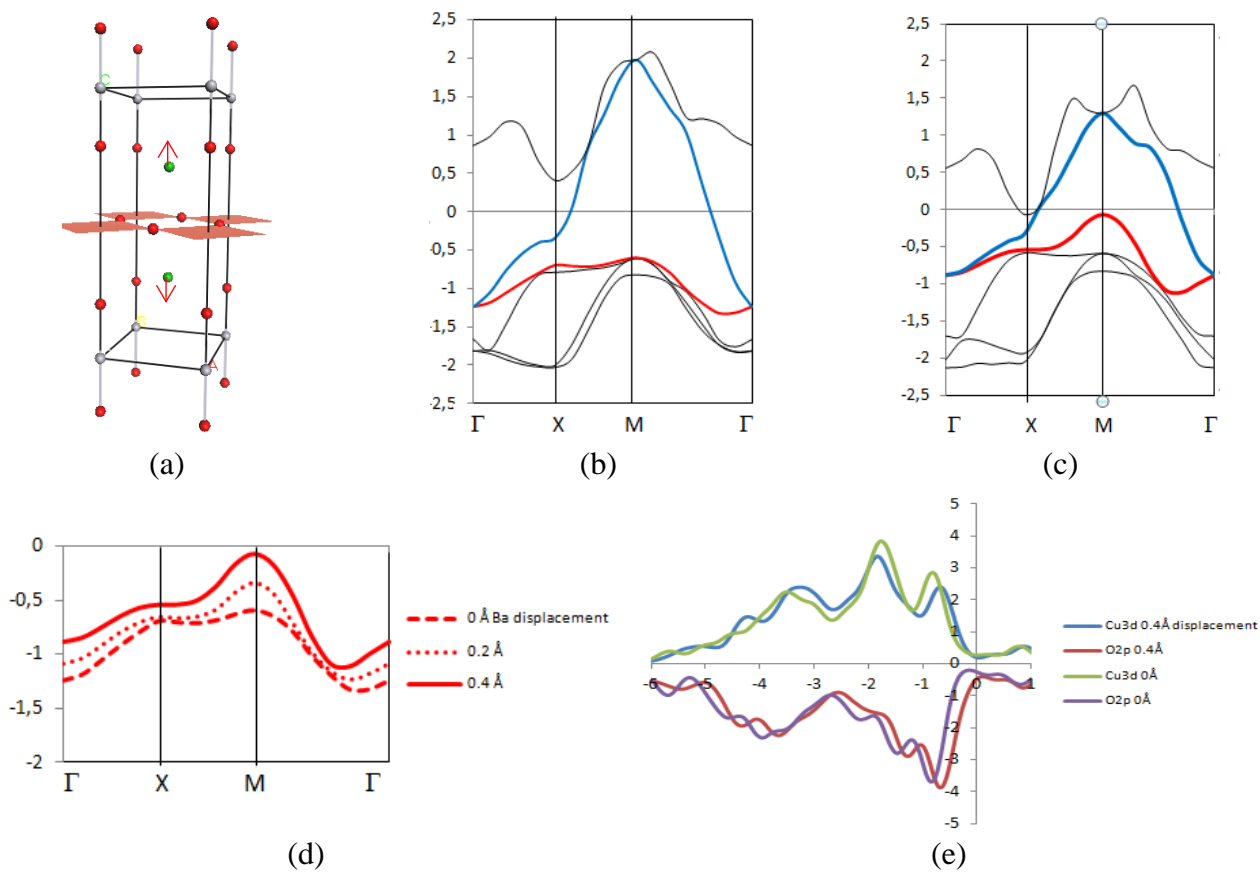

**Figure 1.** *Cont.*

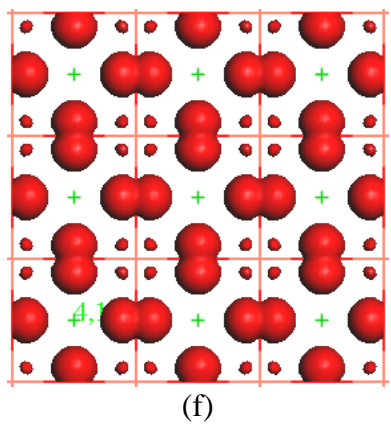

(f)

**Figure 1.** (**a**) The unit cell of HgBa$_2$CuO$_4$ and arrows indicate directions for Ba$^{2+}$ displacements, (**b**) band structure in the vicinity of the Fermi level for experimental strcuture (eV), (**c**) band structure after ±0.2 Å Ba$^{2+}$ displacements (eV), (**d**) effect on band structure of including ±0.1 Å Ba$^{2+}$ displacement (eV), (**e**) the shift of in-plane O2p associated PDOS as compared to effects on Cu3d PDOS (eV), (**f**) The surfacing O2p$_\pi$ band.

Indeed, upon oxygen doping HgBa$_2$CuO$_4$ supercell DFT calculations predict the Ba$^{2+}$ displacements in the vicinity of the added oxygen atom to occur spontaneously, see Figure 2.

**A**

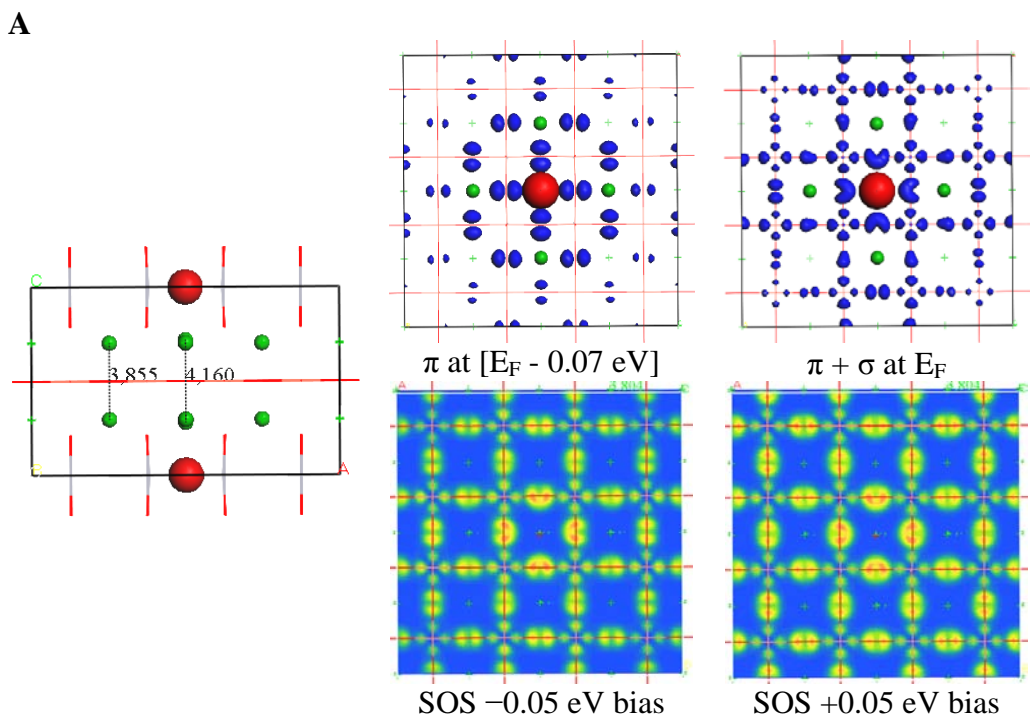

**Figure 2.** *Cont.*

**B**

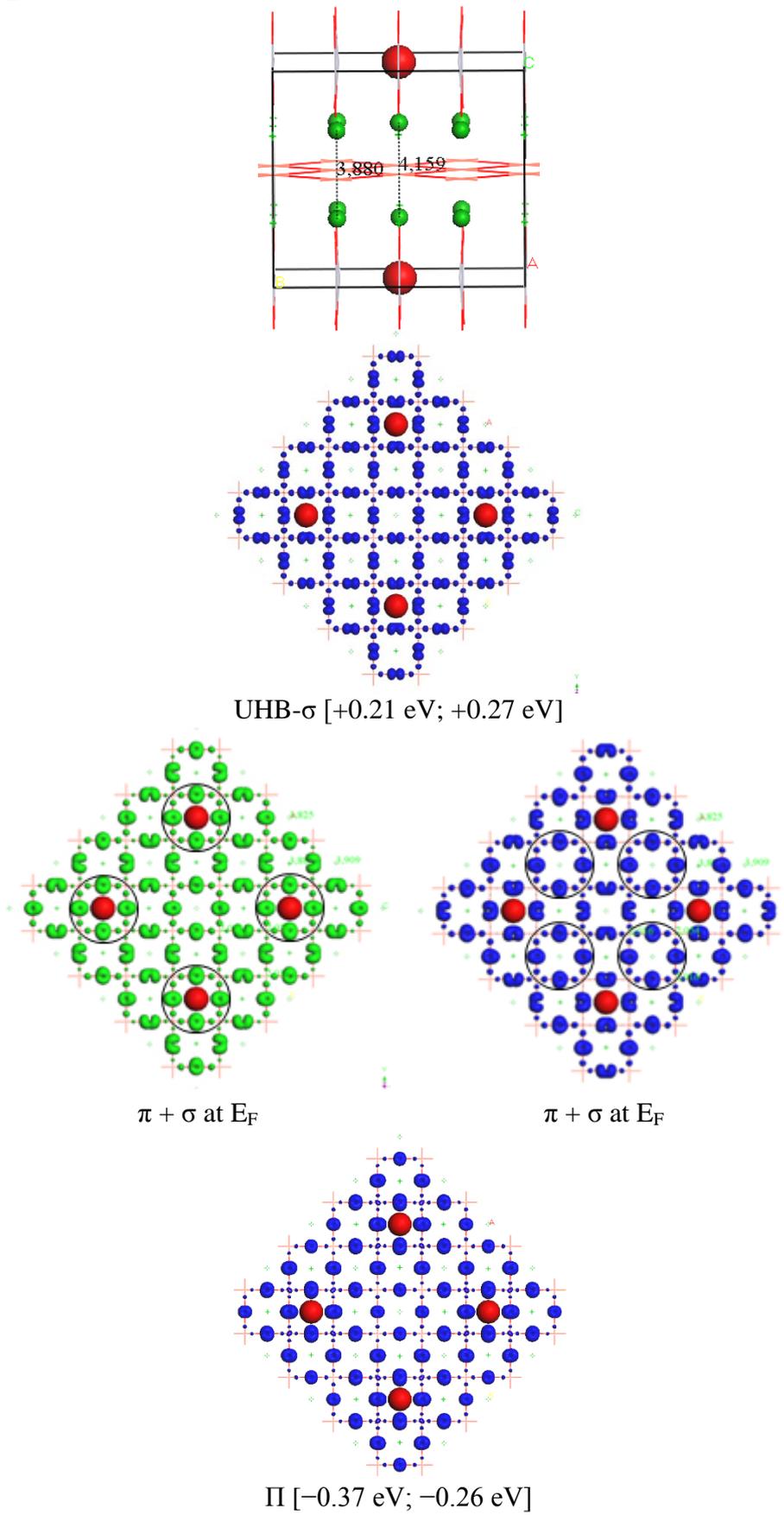

**Figure 2.** *Cont*.

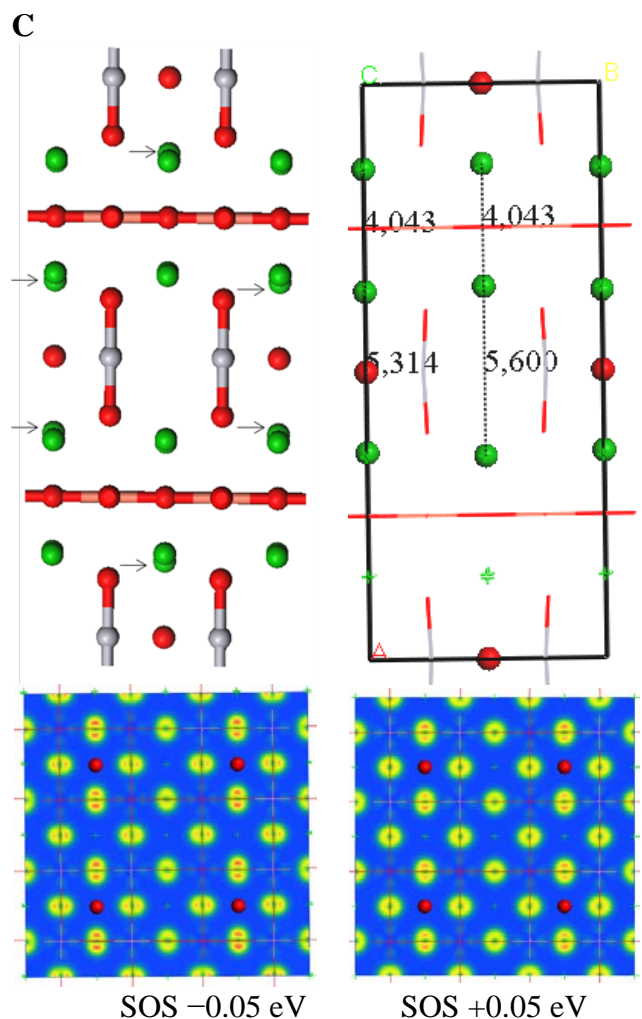

**Figure 2.** Results from supercell calculations are displayed. (**A**) Top: $Hg_{16}Ba_{32}Cu_{16}O_{64+1}$ side view indicating the significant differences in Ba-Ba distances caused by attraction to δO in the Hg plane, left and right panels center: Kohn–Sham states in the vicinity of $E_F$, bottom: −0.05 eV and +0.05 eV bias sum-over-states STM type images in the vicinity of the Fermi level. (**B**) Top: $Hg_8Ba_{16}Cu_8O_{32+1}$ side view again indicating the significant differences in Ba-Ba distances caused by attraction to δO in the Hg plane, left and right panels display the character of crystal orbitals including band energy and dispersion in brackets, in the vicinity of the Fermi level. (**C**) Top: $Hg_8Ba_{16}Cu_8O_{32+2}$ repeatedly the significant differences in Ba-Ba distances as caused by attraction to δO in the Hg plane is observed in agreement with [10]. The doubling of unit cell along the c-axis (2 × 2 × 2) supercell, was in order to test the impact of any artificial oxygen ordering on the electronic structure. Again 0.05 eV and +0.05 eV bias sums-over-states are provided clearly emphasizing the local $O2p_\pi$ character of the states.

As the oxygen atom becomes negatively charged upon hole doping the $CuO_2$ planes, the resulting oxygen ion acts attractor towards the $Ba^{2+}$ ions in its immediate vicinity consistent with experiment [10], see Table 1.

**Table 1.** Ba(N) comprise non-equivalent Ba-to-plane distances. Displacement of proximal Ba(0) towards the δO in the Hg-plane and away from vicinal CuO$_2$ plane. Chu reports estimated displacements *d* of 0.06–0.1 Å from analysis of disorder in neutron powder diffraction studies [10].

| δ = 0 (Figure 1) | Distance to Hg-plane | Distance to CuO$_2$ plane |
|---|---|---|
| Ba(0) | 2.857 Å | 1.893 Å |
| <Ba> [10] | 2.831 Å | 1.919 Å |
| **δ = 0.0625 (Figure 2a)** | **Distance to Hg-plane** | **Distance to CuO$_2$ plane** |
| Ba(0) | 2.670 Å | 2.080 Å |
| Ba(1) | 2.823 Å | 1.928 Å |
| Ba(2) | 2.813 Å | 1.937 Å |
| Ba(3) | 2.830 Å | 1.920 Å |
| Ba(4) | 2.848 Å | 1.902 Å |
| Ba(5) | 2.805 Å | 1.945 Å |
| <Ba> | 2.819 Å (d = 0.15Å) | 1.931 Å |
| **δ = 0.125 (Figure 2b)** | **Distance to Hg-plane** | **Distance to CuO$_2$ plane** |
| Ba(0) | 2.671Å | 2.080 Å |
| Ba(1) | 2.810 Å | 1.940 Å |
| Ba(2) | 2.796 Å | 1.954 Å |
| Ba(3) | 2.838 Å | 1.912 Å |
| <Ba> | 2.793 Å (d = 0.12 Å) | 1.957 Å |
| **δ = 0.25 (Figure 2c)** | **Distance to Hg-plane** | **Distance to CuO$_2$ plane** |
| Ba(0) | 2.657Å | 2.022 Å |
| Ba(1) | 2.800 Å | 2.022 Å |
| Ba(2) | 2.774 Å | 1.971 Å |
| Ba(3) | 2.783 Å | 1.971 Å |
| <Ba> | 2.754 Å (d = 0.10 Å) | 1.997 Å |

The attractive {Ba$^{2+}$ = δO$^-$ = Ba$^{2+}$} structural motif *per se* implies displacement of Ba$^{2+}$ ions away from the CuO$_2$ planes. The observed local surfacing of the *a priori* low-lying O2p(π) band to E$_F$ becomes a manifestation of a change in external potential in line with the Hohenberg–Kohn theorems and is consequently deemed robust. This crystal field attenuations due to reorganization of the ions in the buffer layers and complementary rebuilding of the electronic structure in the CuO$_2$ plane, constitutes one natural cause for the Charge Density Wave (CDW) formations observed in the HTS materials in general and in the Hg cuprates in particular. Emergence of these oxygen "metal" bands and subsequent formation of charge carrier inhomogeneities may well reflect the opening of the so-called pseudogap with complementary recovery of antiferromagnetic (AF) fluctuations. The latter comprise a manifestation of the deconstruction of the electronic structure in the cuprates into two subsystems, *i.e.*, one set of bands related to the AF instability and a disjoint second set of bands associated with the CDW instability. A schematic representation of the rebuilding of the electronic structure due to the dual role of δ-O is provided in Figure 3.

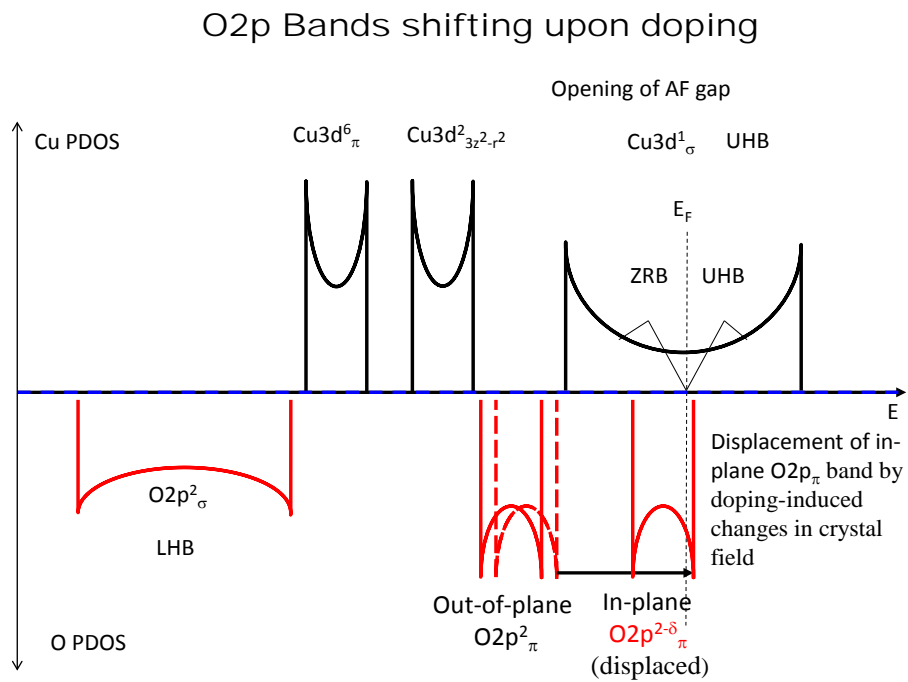

**Figure 3.** The Cu (top) and O (bottom) contributions to the density of states is illustrated schematically. From left to right: At low energies we find the Cu-O2p σ-bonding band, at higher energies emerge the narrow non-bonding Cu3d$_\pi$ bands, the non-bonding O2p$_\pi$ bands emerge at yet higher energies. Dominating at the Fermi level is the Cu contribution to the Cu-O2p σ anti-bonding band, which displays instabilities to antiferromagnetic (AF) gap opening resulting in Zhang–Rice band (ZRB) and upper Hubbard bands (UHB) states bracketing the gap. The present study concerns the surfacing of an O2p$_\pi$ band (dashed) owing to dual role of oxygen doping, *i.e.*, both holes doping and modifying the crystal field acting on the CuO$_2$ planes by pulling the nearby Ba$^{2+}$ ions away from the planes in agreement with experiment [10].

In order to clarify the nature of the new surfacing O2p($\pi$) bands at E$_F$ we acknowledge the CuO$_2$ plane to constitute local quasi-linear Cu-O-Cu coordinations. For a local D$_{2h}$ point group symmetry the non-bonding, bonding and anti-bonding Cu3d($x^2-y^2$)($\sigma$) related orbitals transform in the local A$_g$, B$_{3u}$, and B$_{3u}$ irreducible representations respectively; both bonding and antibonding O2p$_{x,y}$($\sigma$) local orbitals transform in the local B$_{3u}$ irreducible representation, while the two new oxygen orbitals O2p$_{x,y}$($\pi$) transform in the local B$_{2u}$ irreducible representation. The out-of plane O2p$_z$ orbital transforms in the local B$_{1u}$ irreducible representation. It is emphasized that the LHB, ZRB and UHB all belong to the σ category and thus they are protected from mixing/hybridization with the surfacing O2p($\pi$) orbitals. These crucial properties and their possible connection to D-wave origin of HTS in the cuprates is provided in Figure 4.

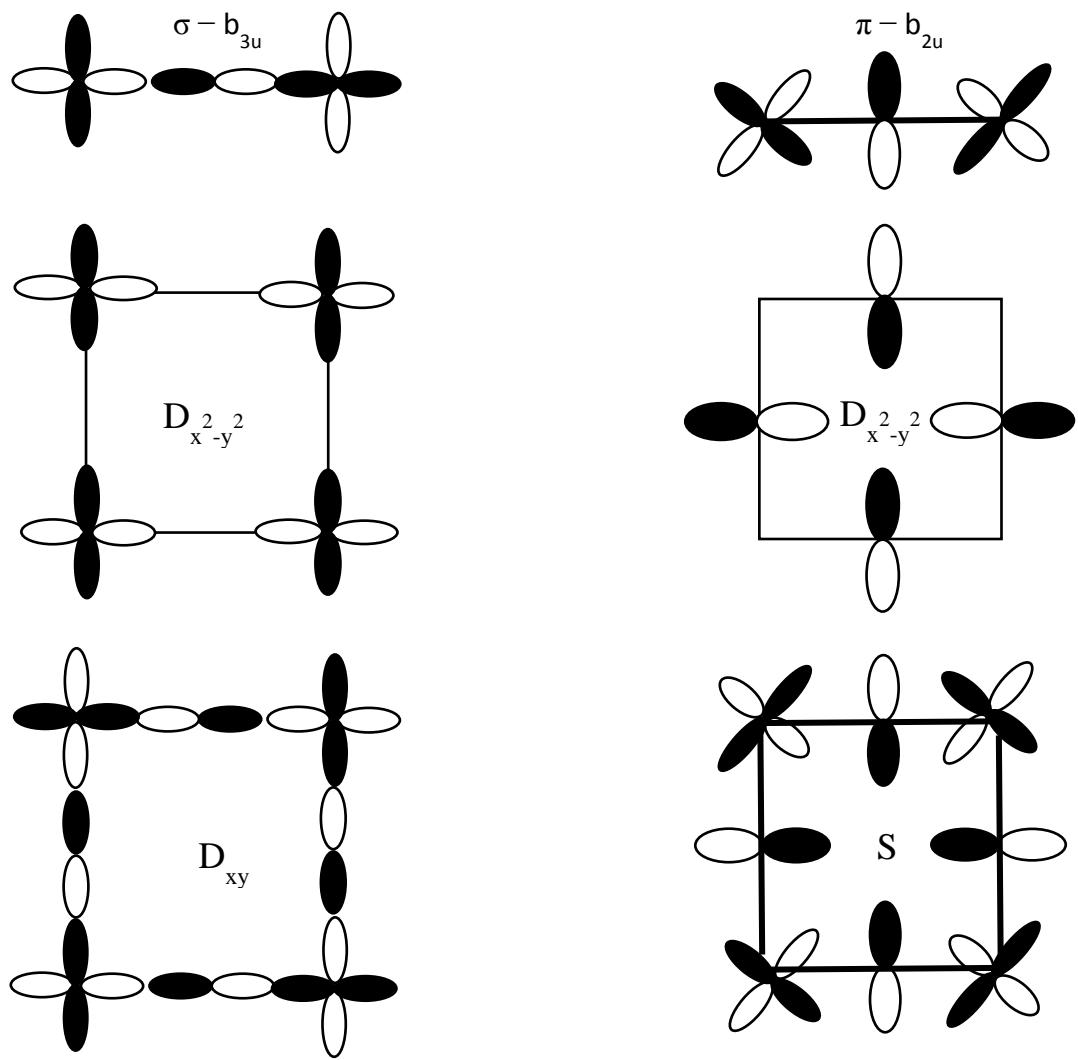

**Figure 4.** Left column. Top: Local orbital signature of upper Hubbard bands (UHB) and Zhang–Rice band (ZRB), belonging to the $B_{3u}$ irreducible representation. Center: Node-less linear combination of Cu3d($x^2-y^2$)($\sigma$) orbitals of "D" symmetry. Bottom: Linear combination of Cu3d($x^2-y^2$)($\sigma$) and O2p$_{x,y}$($\sigma$) orbitals with two nodal planes of "D" symmetry. Right column. Top: Local orbital signature of $\pi$ band of belonging to the $B_{2u}$ irreducible representation. Center: Linear combination of O2p$_{x,y}$($\sigma$) orbitals with two nodal planes of "D" symmetry. Bottom: Node-less linear combination involving O2p$_{x,y}$($\pi$) and Cu3d(xy) orbitals of "S" symmetry.

The Cu (top) and O (bottom) contributions to the density of states is illustrated schematically. From left to right: At low energies we find the Cu-O2p $\sigma$-bonding band, at higher energies emerge the narrow non-bonding Cu3d$_\pi$ bands, the non-bonding O2p$_\pi$ bands emerge at yet higher energies. Dominating at the Fermi level is the Cu contribution to the Cu-O2p $\sigma$ anti-bonding band, which displays instabilities to AF gap opening resulting in ZRB and UHB states bracketing the gap. The present study concerns the surfacing of an O2p$_\pi$ band (dashed) owing to dual role of oxygen doping, *i.e.*, both holes doping and modifying the crystal field acting on the CuO$_2$ planes by pulling the nearby Ba$^{2+}$ ions away from the planes in agreement with experiment [10].

The possibly crucial importance of these finding is in the coexistence of the σ and π bands at a common chemical potential providing support to understandings of HTS as an effective two-bands/two-gaps phenomenon, *i.e.*, the opening of an antiferromagnetic AF gap in σ is related to a complementary opening of a charge density wave gap in π, *cf.* Figure 5.

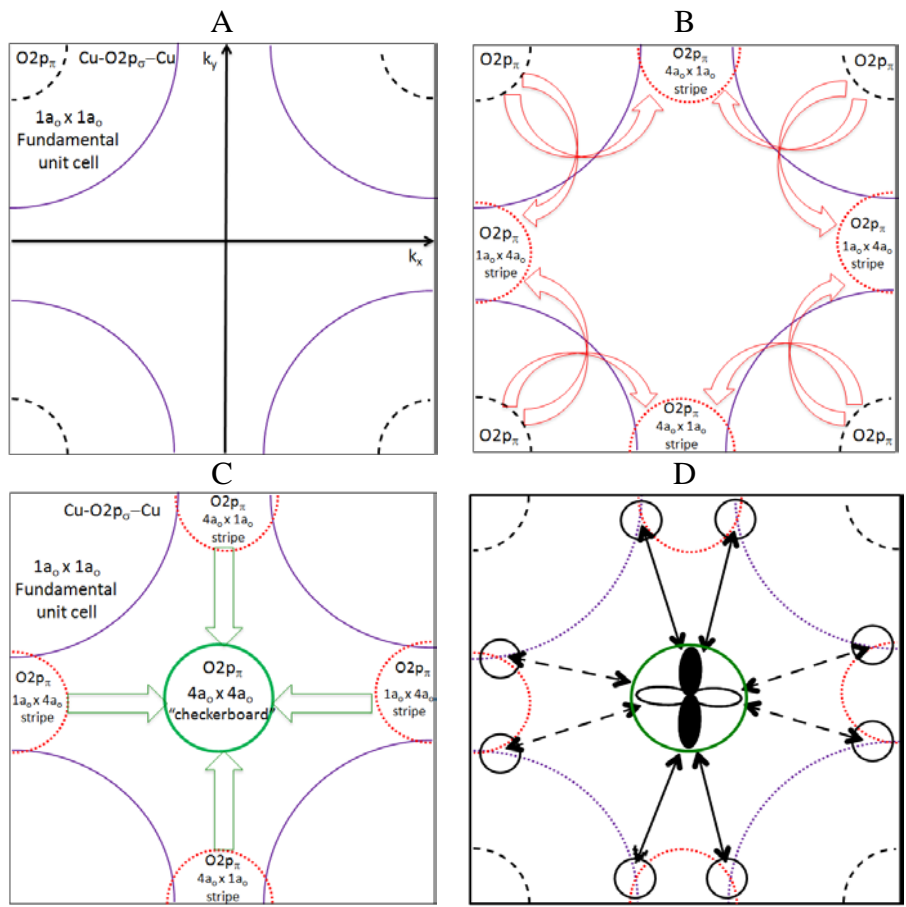

**Figure 5.** (**A**) Bands of $O2p_\pi$ and $Cu-O2p_\sigma-Cu$ origins cross the common Fermi energy caused by crystal field induced transfer of charge carriers. (**B**) Folding of $O2p_\pi$ hole pockets at M point due to X, and Y centered stripes instabilities. (**C**) Folding of stripes superstructures onto checkerboard superstructure. (**D**) Coupling of locally orthogonal π and σ states at octet of hot-spots indicated, being due to symmetry violation upon formation of superstructures. *cf.* $D(x^2-y^2)$ symmetry spanned by σ (Figure 4 left) and π (Figure 4 right) local states. The resulting many-body order parameter of D-wave symmetry is indicated, *cf.* [5,11].

Were this to hold, then any theoretical effort attempting to model the superconductivity in the hole doped cuprates while employing the band structure of the non-doped cuprates, should be deemed inadequate. This is not because of the shortcomings of DFT but due to the lack of relevance of the non-doped $HgBa_2CuO_4$ for describing the normal state of the doped $HgBa_2CuO_{4+\delta}$. This is consistent with the findings in [12]. In contrast, usefulness of explicit supercell DFT calculations to address the impacts of oxygen dopants was demonstrated above.

## 2. Results and Discussion

In what follows the relevance to HTS of two bands coexisting at a common Fermi energy is discussed. In doing so, a cooperative electronic "gas-solid" instability is envisaged. At 0 K the two electronic subsystems display quantum analogs of two grand canonical ensembles, *i.e.*, each acting "bath" for the other with respect to virtual quantum fluctuations [13–15]. The total energy and all properties commuting with the total Hamilton operator are conserved. The inter-subsystem non-adiabaticity implies virtual quasiparticle excitations in one subsystem being absorbed by corresponding virtual quasiparticle excitations in the other subsystem, *i.e.*, virtual double excitations composed of coupled virtual single-excitation contributing to the correlated ground state. While both electronic systems are gaped when decoupled, inter-subsystem coupling causes virtual partial evaporations in the two by the coupled single quasi-particle excitations across each of the two gaps. Thus the two subsystems cooperatively mix insulator and metal properties. In doing so they satisfy the requirements of superconductivity, *i.e.*, said coupled virtual single excitations offer a generalization of the Bardeen–Cooper–Schrieffer (BCS) virtual double excitations contributing to the BCS 0 K wave function [16]. To clarify this further, consider again:

$$|\psi_{BCS}\rangle = \prod_{-\chi<k<\chi} \left( u_{k_F-k} |vac\rangle_{k_F-k} + v_{k_F-k} |\uparrow\downarrow\rangle_{k_F-k} \right) \tag{1}$$

reflecting the degeneracy at the Fermi level of an electron gas normal state and the effect of non-adiabatic electron-phonon coupling. When assuming particle-hole symmetry Equation (1) transforms into:

$$|\psi_{BCS}\rangle = \prod_{k<\chi} \left( v_{k_F-k} |\uparrow\downarrow\rangle_{k_F-k} + v_{k_F+k} |\uparrow\downarrow\rangle_{k_F+k} \right) \tag{2}$$

implying correlation of pair-states symmetrically above and below $k_F$ by means of virtual double-excitations from $k_F - k$ to $k_F + k$ such that $v_{k_F+k}^2 + v_{k_F-k}^2 = 1$. A Peierls type CDW insulator instability becomes an implicit property of Equation (2), *i.e.*,:

$$|\psi_{BCS}\rangle = \prod_{k<\chi} \left( v_{k_F-k}^g |\uparrow\downarrow\rangle_{k_F-k}^g + v_{k_F-k}^u |\uparrow\downarrow\rangle_{k_F-k}^u \right) \tag{3}$$

where $g$ is gerade and $u$ is ungerade, referring to the lower and upper bands in the resulting reduced Brillouin zone. This Fermi surface nesting instability is employed in the present study to extend on the BCS approach. Thus we envisage two nesting instabilities, AF and CDW, at the common Fermi energy. It is shown explicitly in Figure 4 how the same $D_{x^2-y^2}$ local orbital symmetry may result from dual nesting at the common Fermi surface, where the nesting of one benefits from the nesting of the other. Superconductivity owing to electron correlation on an effective Fermi surface, *cf.* Figure 5, and entanglement of AF and CDW subsystems, *cf.* arrows in Figure 5d correlating the octet of hot-spots to the checkerboard CDW superstructure (*cf.* e.g., [5,11]), implies correlated wave functions on the form:

$$\left|\psi_{CDW}\right\rangle \otimes \left|\psi_{AF}\right\rangle = \prod_{\substack{\left|E^{CDW}_{k_F-k} - E^{CDW}_F\right| \leq \varepsilon \\ \left|E^{AF}_{k'_F-k'} - E^{AF}_F\right| \leq \varepsilon}} \left\{ \theta^{k'_F-k'}_{k_F-k} \left[ \left|P^{k_F-k}_{CDW}\right\rangle \left|S^{k'_F-k'}_{AF}\right\rangle + \left|B^{k_F-k}_{CDW}\right\rangle \left|T^{k'_F-k'}_{AF}\right\rangle \right] + \gamma^{k'_F-k'}_{k_F-k} \left|A^{k_F-k}_{CDW}\right\rangle \left|D^{k'_F-k}_{AF}\right\rangle \right\} \quad (4)$$

where quasi-molecular type correlated pair and pair-broken states in the charge density wave channel comprise:

$$\left|P^{k_F-k}_{CDW}\right\rangle = p_{k_F-k}\left( v^g_{k_F-k} \left|\uparrow\downarrow\right\rangle^g_{k_F-k} + v^u_{k_F-k} \left|\uparrow\downarrow\right\rangle^u_{k_F-k} \right) \quad (5a)$$

$$\left|B^{k_F-k}_{CDW}\right\rangle = b^{ug}_{k_F-k} \left|\uparrow\right\rangle^g_{k_F-k} \times \left|\uparrow\right\rangle^u_{-(k_F-k)} \quad (5b)$$

$$\left( v^g_{k_F-k} \right)^2 + \left( v^u_{k_F-k} \right)^2 = 1 \quad (5c)$$

and similarly, pair and pair-broken states in periodic quasi-molecular analog to the AF channel comprise:

$$\left|S^{k_F-k}_{AF}\right\rangle = s_{k'_F-k'}\left( w^g_{k'_F-k'} \left|\uparrow\downarrow\right\rangle^g_{k'_F-k'} + w^u_{k'_F-k'} \left|\uparrow\downarrow\right\rangle^u_{k'_F-k'} \right) \quad (6a)$$

$$\left|T^{k'_F-k'}_{AF}\right\rangle = t^{ug}_{k'_F-k'} \left|\downarrow\right\rangle^g_{k'_F-k'} \times \left|\downarrow\right\rangle^u_{-(k'_F-k')} \quad (6b)$$

$$\left( w^g_{k'_F-k'} \right)^2 + \left( w^u_{k'_F-k'} \right)^2 = 1 \quad (6c)$$

Said requirement of the two sub-systems being complementary quantum grand canonical ensembles implies:

$$p_{k_F-k} = s_{k'_F-k'} \equiv pq^{k'_F-k'}_{k_F-k} \quad (7a)$$

$$b^{ug}_{k_F-k} = t^{ug}_{k'_F-k'} \equiv bt^{k'_F-k'}_{k_F-k} \quad (7b)$$

$$\left( pq^{k'_F-k'}_{k_F-k} \right)^2 + \left( bt^{k'_F-k'}_{k_F-k} \right)^2 = 1 \quad (7c)$$

but also the inclusion of contributions from inter-subsystem virtual double-excitations:

$$\left|A^{k_F-k}_{CDW}\right\rangle \left|D^{k'_F-k}_{AF}\right\rangle = \left|\uparrow\downarrow\right\rangle^{g,CDW}_{k_F-k} \times \left|\uparrow\downarrow\right\rangle^{u,CDW}_{k_F-k} \times \left|vac\right\rangle^{AF}_{k'_F-k'} \quad (8)$$

where the CDW band acts electrons acceptor and the AF band acts electron donor, while the global number of particles is a conserved quantity. By virtue of Equations (4) and (7) the overall normalization requirement can be written:

$$\left( \theta^{k'_F-k'}_{k_F-k} \right)^2 + \left( \gamma^{k'_F-k'}_{k_F-k} \right)^2 = 1 \quad (9)$$

It is crucially noted that the inter-band virtual double excitation contributions are possible only when the two subsystems display the same chemical potential *i.e.*, have a common Fermi energy:

$$E_F^{CDW}(Q_{CDW}) = E_F^{AF}(Q_{AF}) \tag{10}$$

where $Q_{CDW}$ and $Q_{AF}$ refer to the two corresponding superstructures. We observe that Equation (4) represents an analog to BCS theory for two coupled "electron gases" where Equation (8) reflects a complementary Josephson term, responsible for the superfluidity owing to quantum CDW fluctuations. In particular, Equation (4) states that embedded in the superconducting phase is a dual Quantum Critical Point (QCP) composed of:

$$\{antiferromagnet \leftrightarrow quantum\ paramagnet\}(\sigma_{QCP})$$

and:

$$\{"strange\ metal"(CDW) \leftrightarrow undoped\ insulator\}(\pi_{QCP})$$

one acting embedding for the other, implying superconductivity to emerge from cooperation rather than competition [13–15]. A characteristic feature of the σ-π understanding is that *particle–hole* symmetry is violated. This is because of the different energy dispersions of the AF(σ) and CDW(π) channels which jointly accommodate the composite Cooper pairs, in line with experimental observations and previous work [9,17].

The inter-band Josephson term in Equation (4) may be generalized to include inter-plane coupling owing to the degeneracy among equivalent planes:

$$\{|\psi_{CDW}\rangle \otimes |\psi_{AF}\rangle\}_1 \{|\psi_{CDW}\rangle \otimes |\psi_{AF}\rangle\}_2 \{|\psi_{CDW}\rangle \otimes |\psi_{AF}\rangle\}_3 \ldots$$

which is how three-dimensional superconductivity develops in the hole doped cuprates. The enhanced superconductivity in double-plane Hg1212 becomes a special case resulting from proximity between equivalent planes. The impact of introducing intervening $CuO_2$ planes, *i.e.*, one in case of Hg1223, and two in case of Hg1234 can be appreciated from supercell DFT, *cf.* Figures 6.

Again δ-O induced displacement of adjacent $Ba^{2+}$ ions is observed. It is noted however, that the different crystal fields acting on inner and outer $CuO_2$ planes cause the UHB of the inner $CuO_2$ plane to come close to the common Fermi energy. The gap between ZRB (outer) and UHB (inner) is still significant in case of Hg1223 causing a negligible or even positive effect on the superconductivity. For definitions, see again Figure 3. Said gap is reduced significantly for Hg1234 effectively resulting in the injection of additional holes into outer $CuO_2$ planes. In as much as Hg1201, and Hg1212 are underdoped, it is maybe surprising to find the calculations to imply introduction of intervening $CuO_2$ planes to cause increased hole doping only, and possibly to the extent that even the overdoped regime may be accessed upon increasing the number of such planes. The approximate cylinder symmetry reflecting O2p(σ) mixing with O2p(π) comes out clearly in outer $CuO_2$ planes, while O2p(σ) dominates for inner $CuO_2$ planes reflecting their UHB character. It is finally noted that this doping effect does not display the dual role of δ-O and that the pressure effect may result from a combination of the two.

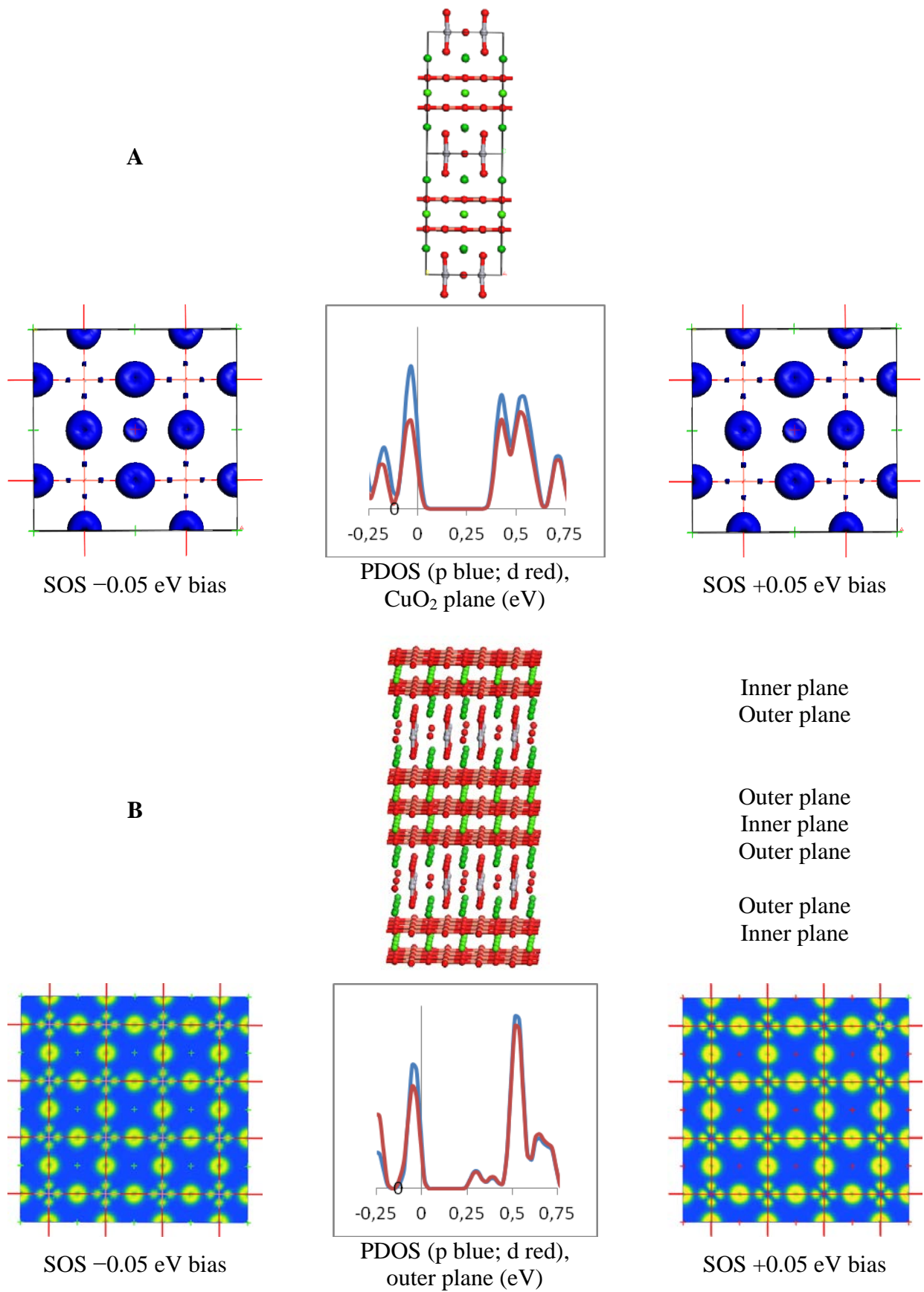

Figure 6. *Cont*.

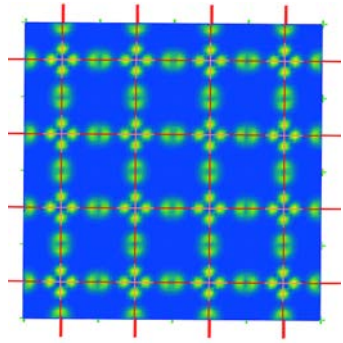

SOS −0.05 eV bias

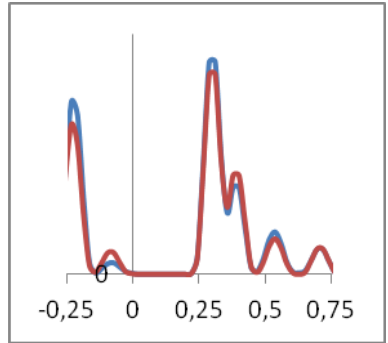

PDOS (p blue; d red), Inner plane (eV)

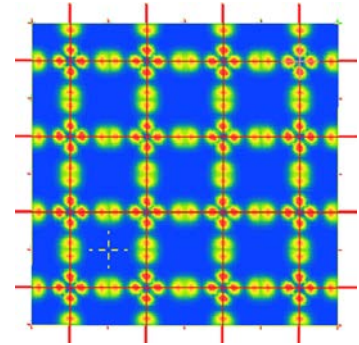

SOS +0.05 eV bias

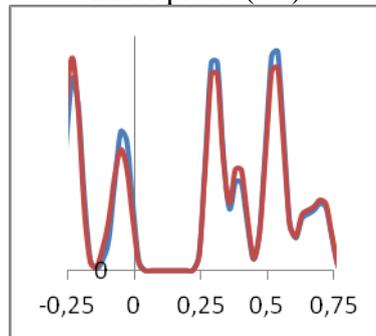

PDOS (p blue; d red), sum of Inner and Outer planes (eV)

**C**

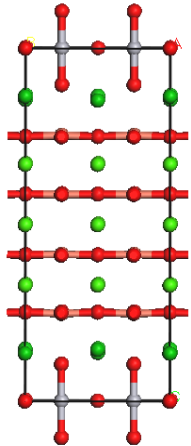

Outer plane
Inner plane
Inner plane
Outer plane

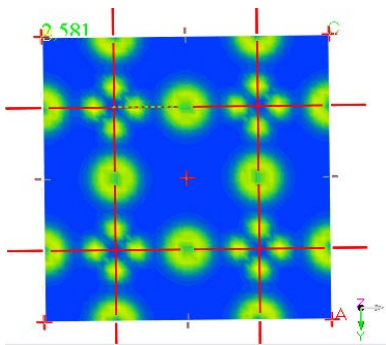

SOS −0.05 eV bias

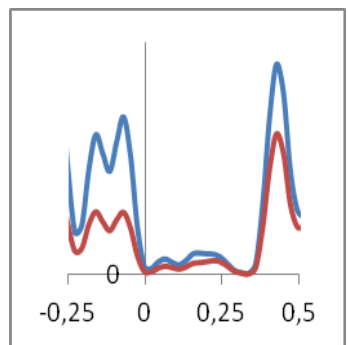

PDOS (p blue; d red), outer plane (eV)

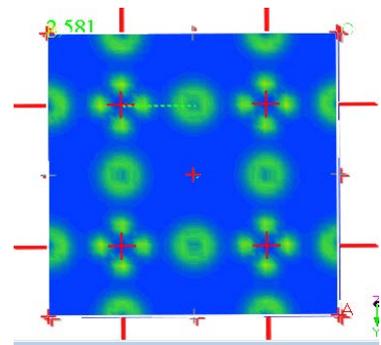

SOS +0.05 eV bias

**Figure 6.** *Cont*.

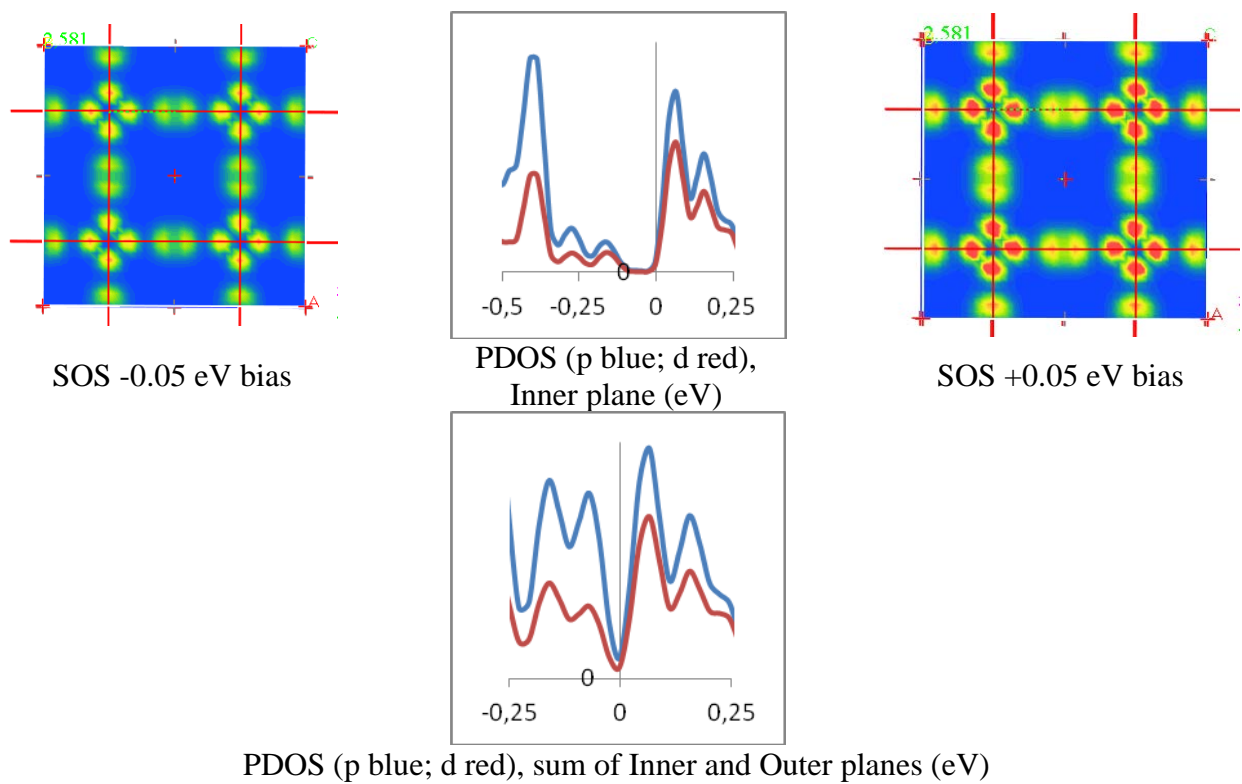

**Figure 6.** Results from supercell calculations on (**A**) "Hg1212", here $Hg_4Ba_8Ca_4Cu_8O_{24+1}$, (**B**) "Hg1223", here $Hg_4Ba_8Ca_8Cu_{12}O_{32+1}$, and (**C**) "Hg1234", here $Hg_4Ba_8Ca_{12}Cu_{16}O_{40+1}$, are displayed. In (A) 1st row displays the crystal structure, 2nd row: left is sum over states SOS −0.05 eV bias; right is SOS +0.05 eV bias; middle is partial density of states (blue p states, red d-states). In each of (B) and (C) 1st row displaying the crystal structures. 2nd row is Outer Plane: left is sum over states SOS −0.05 eV bias; right is SOS +0.05 eV bias; middle is partial density of states (blue p states, red d-states). 3rd row is Inner Plane: left is sum over states SOS −0.05 eV bias; right is SOS +0.05 eV bias; middle is partial density of states (blue p states, red d-states). 4th row is the sum of PDOS for inner and outer planes. Negligible impact of central plane on total PDOS at $E_F$ for Hg1223 is contrasted by the result for Hg1234. Approximate cylinder symmetry reflecting O2p(σ) mixing with O2p(π) comes out clearly.

## 4. Conclusions

What renders the superconducting state stable at high temperatures is the ambiguity in ways to satisfy the global quantum mechanical constraints accessible to the real system in its ground state, *i.e.*, by inter-subsystem entangled coupled virtual quasiparticle excitations, *cf.* Equations (5) and (6), the sharing of Cooper pairs, *cf.* Equation (8) in the context of Equation (4), thus adding elasticity to the brittle AF + CDW correlations. Below the critical temperature for superconductivity the virtual states contributing to the correlated ground state constitute a von Neumann entropy, which is cancelled by the entanglement of said virtual states. At finite temperatures thermal quasiparticle excitations across the pseudogap (PG) and/or spingap (SG) access the very same portions of Hilbert space, albeit decoupled, as are accessed by the entangled many-body wave function of superconductivity. Building on this competition, any strategy to improve the stability of the composite Cooper pairs must reconcile the opening of gaps for quasiparticle excitations across the PG/SG while maintaining near-degeneracy among the ensemble of virtual two-channel many-body states contributing to the correlated ground state.

The superconducting dome results from two antagonistic contributions: (a) oxygen doping increases the Hilbert space of virtual states accessible by the electron correlation (competing with the thermal excitations across either of the two single particle gaps rendering said states real), and (b) the overstepping of the largest Q-value of a viable CDW instability upon overdoping.

Irrespective of whether the CDW is of checkerboard form [9,17] or it is striped, the superconducting order parameter represents the matching of the virtual coupled single-excitations Equations (5b) and (6b) in Equation (4). Protection against intra-band chemical bonding is obtained by utilizing the non-bonding atomic O2p($\pi$) orbitals. Inter-band chemical bonding is avoided because the $\sigma$ and $\pi$ bands belong to different local irreducible representations. The latter avoids in particular the Zhang–Rice singlet pairing, which is detrimental to the superconductivity. Indeed, upper bound for the superconducting gap is reflected in the complete recovery of superexchange interaction of the undoped $\sigma$-bands upon inter-band $\sigma - \pi$ hole transfer.

Superconductivity in the holes doped $CuO_2$ planes may emerge from a normal state possessing two mutually protected small-gaped bands at a common Fermi energy. Besides us, the potential importance of O2p($\pi$) associated bands for superconductivity in the holes doped cuprates has previously been proposed mainly by Goddard [18], Hirsch [19], and recently also from analysis of angle resolved photoemission spectroscopy [20]. The requirement of bands from orbitals belonging to different irreducible representations coalescing at $E_F$ renders high-$T_C$ superconductivity rare in nature. The demonstrated dual role of oxygen in $HgBa_2CuO_{4+\delta}$ offering simultaneous holes doping and crystal field tuning [21] is promising because: (a) it implies that the entangled instabilities can be deliberately tuned and (b) it show how present days supercell DFT can be utilized in search for materials that display superconductivity possibly also at room temperature.

## 5. Computational Details

DFT subject to the General Gradient Approximation (GGA) utilizing the Perdew Burke Ernzerhof (PBE) functional [22] in conjunction with supercell calculations employing ultrasoft pseudo-potential projectors computed "on the fly" [23] as implemented in [24] including a 550 eV cut-off energy were employed throughout. Convergence criteria included $10^{-5}$ eV/atom for total energy, 30 meV/Å for force, and 0.1 pm for distance. Additional information is included in the corresponding figure captions. The k-point sets reflect the sizes of the supercells, *i.e.*, ($3 \times 3 \times 2$) sets were employed in case of Hg1212+$\delta$, Hg1223+$\delta$, and Hg1234 + $\delta$. In case of Hg1201+$\delta$ the chemical unit cell calculations utilized a ($6 \times 6 \times 3$) set, while the ($2 \times 2 \times 3$) set was employed for most Hg1201+$\delta$ supercell calculations besides the $Hg_{16}Ba_{32}Cu_{16}O_{64+1}$ for which gamma point calculation was deemed sufficient.

**Acknowledgments**

Over the years I have benefited greatly from discussions with Tord Claeson. Alexander Föhlisch, Michael Odelius, Justine Schlappa, Piter Sybren, Martin Beye are acknowledged for work in progress at BESSY II—how validate the possible presence of states of local σ and π symmetry coalescing at $E_F$ in Hg1201 by means of angle resolved Resonant Inelastic X-ray Scattering RIXS. Support from the Swedish Research Council, grant "Breakthrough Research", is gratefully acknowledged.